\documentclass{ws-ijmpd}
\usepackage[super,compress]{cite}
\usepackage[dvips,breaklinks]{hyperref}
\hypersetup{colorlinks,urlcolor=black,citecolor=black,linkcolor=black,filecolor=black}
\usepackage{breakurl}
\usepackage{}
\begin{document}


%
\catchline{}{}{}{}{}
%

\title{Shear Dynamics in Higher Dimensional FLRW Cosmology}

\author{Isha Pahwa}

\address{Department of Physics and Astrophysics, University of Delhi,\\
Delhi  $110007$,
India\\
ipahwa@physics.du.ac.in}

\author{Hemwati Nandan}

\address{Department of Physics, Gurukula Kangri Vishwavidyalaya, \\
Haridwar – $249404$, Uttarakhand, India\\
hnandan@iucaa.ernet.in}

\author{Umananda Dev Goswami}
\address{Department of Physics, Dibrugarh University,\\
Dibrugarh $786004$, Assam, India\\
umananda2@gmail.com}
\maketitle

\begin{history}
\received{Day Month Year}
\revised{Day Month Year}
\end{history}

\begin{abstract}
We study the shear dynamics of higher dimensional 
Friedmann-Lema\^{i}tre-Robertson-Walker (FLRW) cosmology by considering a
non-perfect fluid which exerts different pressure in the 
normal and extra dimensions. We generalise the definition of shear tensor for 
higher dimensional space-time and prove it to be 
consistent with the evolution equation for shear tensor obtained from the 
Ricci identities. The evolution of  shear tensor is investigated numerically. 
The role of extra dimensions and other parameters involved in shear dynamics 
is discussed in detail. We find that with increase in anisotropy parameter, 
time of decay of shear increases while with increase in number of 
extra dimensions, shear tends to decay early.
\end{abstract}

\keywords{Extra dimensions, FLRW cosmology, shear and non-perfect fluids}

\ccode{PACS numbers: 98.80.Jk, 04.50.Cd}


\section{Introduction}
Starting with their introduction by Kaluza \cite{Kaluza} and Klein \cite{Klein},
various models in gravity and cosmology in the context of higher  
dimensions have been used in recent times , where the matter fields reside in all 
dimensions including the compact extra dimensions. Such models have received a considerable
support from string theory which provides a strong mathematical basis for the description 
of the universe in terms of higher dimensions. Further, there is a possibility that these models in the context of 
particle phenomenology and cosmology could provide a natural explanation to the well-known 
issues like hierarchy problem. Another motivation which is more fundamental to cosmology 
is that extra dimensions may play an important role in early universe.
In the early phase, the universe could have been described by the `brane world models' 
in which the standard model fields are trapped on a three dimensional 
hypersurface (i.e. brane), embedded in a higher dimensional spacetime \cite{tasi}. 
There are two such popular models, viz., 
the Arkani-Dimopolous-Dvali (ADD) model \cite{ADD} and Randall-Sundrum (RS) 
model \cite{RS1, RS2}, which have testable predictions at 
present day colliders.

In the early universe, the energy of the universe was typically high enough to make 
the existence of very small extra dimensions perceptible. 
The dynamics of the universe could have been different as compared to the normal 
$(1+3)$ dimensional case due to the presence of the scale factor of extra dimensions.
The scale factor of extra dimensions will in general be different from that
of the normal dimensions. This motivates us to study the evolution of the 
universe with the extra dimension in the context of cosmology. 
For this purpose, one may study the 
kinematical quantities such as expansion scalar, shear and rotation (or 
vorticity) in order to see how the dynamics of the universe evolves with 
different conditions on scale factors as well as on the number of extra 
dimensions. The expansion scalar determines the rate of change of distance of 
neighbouring  particles in the fluid (which is related to the volume expansion),
while the shear tensor determines the distortions arising in the fluid flow keeping
the volume preserved. The vorticity tensor however determines a rigid rotation 
of clusters of galaxies with respect to a local interial rest frame. It would 
be interesting to investigate the  evolution of such quantities with 
extra dimensions. Our main objective here is to discuss the shear dynamics of 
the universe in higher dimensions with the 
Friedmann-Lema\^{i}tre-Robertson-Walker (FLRW) cosmology. 

The other motivation to do this analysis comes from an unresolved issue 
which is the effect of the structure formation on the cosmological 
observation beyond perturbative analysis of the FLRW models \cite{mattsson}.  
The effect on the cosmic structures may be estimated via a backreaction term 
that 
arises by averaging inhomogeneous scalar quantities on spatial hypersurfaces. 
Thus, the backreaction term is calculated by subtracting the non-negative 
average shear from expansion rate. It motivates us to look at scenarios other than
FRW models.

The paper is organised as follows. In the next section, we will present the 
necessary formalism to introduce the definition of the concerned kinematical 
quantities namely the expansion scalar, shear and rotation tensors to describe 
the behaviour of the geodesic congruence along with the Raychaudhuri equation 
for a congruence of timelike geodesics. In Section $3$, we develop the shear 
dynamics of FLRW cosmology with extra spatial dimensions by calculating the 
non-vanishing components of shear tensor in view of the solutions of scale 
factors as obtained from the Einstein equations. We present the evolution of 
the expansion scalar and components of shear numerically in Section $4$ for 
different number of extra dimensions. In the last section, we summarise our 
results with future possibilities.

\section{$1+n$ Decomposition and Kinematical Quantities}
We introduce a family of observers with non-intersecting world lines. The 
timelike $4-$velocity vector which is tangent to these world lines is
$u^{\alpha}=\frac{dx^{\alpha}}{d\tau}$ 
and satisfies $u_{\alpha} u^{\alpha} =-1$, where $\tau$ is proper time 
measured along the world lines. The spacetime metric can be 
expressed  the longitudinal  $(-u_{\alpha} u_{\beta})$ and transverse 
parts ($h_{\alpha \beta}$) as,

\begin{equation}
h_{\alpha \beta} = g_{\alpha \beta} + u_{\alpha} u_{\beta}.
\end{equation}
The latter projects the spacetime orthogonal to the $4-$velocity into the 
observers' instantaneous rest space at each event. The velocity of a comoving 
particle is given as $u_{\alpha} = (1,0,0,0)$. 

The vector field $u_{\alpha}$ and its associated tensor counterpart 
$h_{\alpha \beta} $ allows for a unique
decomposition  of every spacetime quantity into its irreducible timelike and
spacelike parts ~\cite{ellis1999,tsagas}. In order to characterise the 
evolution of the world lines, the covariant derivative of the velocity field may be split 
into its irreducible parts, defined by their symmetric properties:

\begin{equation}
 u_{\beta;\alpha} = -u_{\alpha} \dot{u_{\beta}} + \frac{1}{n} \Theta h_{\alpha \beta} + \sigma_{\alpha \beta} + \omega_{\alpha \beta}.
\end{equation}
Here, the trace part $\Theta=u^{\alpha}_{;\alpha}$ denotes the rate of 
expansion of the fluid, the symmetric-tracefree part $\sigma_{\alpha \beta} = 
(1/2)(u_{\beta;\alpha} + u_{\alpha;\beta})$ is the shear tensor and the 
anti-symmetric part $\omega_{\alpha \beta}= (1/2)(u_{\alpha;\beta} - u_{\beta;
\alpha})$ is the rotation tensor. Here, $n$ is the total number of spatial 
dimensions. The governing equations of these quantities 
are the propagation equations \cite{dadhich,ellis2007,kar2007} which give the 
overall evolution (along the flow) of these quantities, i.e., $\Theta, 
\sigma_{\alpha \beta} $ and $\omega_{\alpha \beta}$ in a given background 
spacetime. In particular, the evolution of $\Theta$ is governed by 
the Raychaudhuri equation given by:
\begin{equation}
 \frac{d\Theta}{d\tau} = -\frac{1}{n} \Theta^2 - \sigma^{\alpha \beta}\sigma_{\alpha \beta}\, + \, \omega^{\alpha \beta}\omega_{\alpha \beta}\, - \, R_{\alpha \beta} \, u^{\alpha} u^{\beta} \label{rayeqn}
\end{equation}
We assume the universe to be filled with a fluid which is irrotational as 
otherwise the velocity will  die down rapidly with expansion of universe. The 
tangent planes of the comoving observers together form spacelike 
hypersurfaces which are normal to the world lines of the observers. More 
precisely, we consider the geodesic congruences which have rotation tensor, 
$\omega_{\alpha \beta} = 0$ as asserted by the {\it Frobenius' Theorem}. The 
projected Riemann tensor \cite{tsagas} on the hypersurface is defined 
as 
\begin{equation}
 \mathcal{R}_{abcd} = h_a^q \, h_b^s \, h_c^f \, h_d^p \, R_{qsfp} \, - \,  u_{a;c} \, u_{b;d} + u_{a;d} \, u_{b;c} 
\end{equation}
The local Ricci tensor $\mathcal{R}_{ab}$ and Ricci scalar $\mathcal{R}$ on 
the hypersurface orthogonal to $u_a$ are defined by
\begin{equation}
 \mathcal{R}_{ab} = h^{cd} \, \mathcal{R}_{cadb} \quad \quad \mathcal{R} = h^{ab} \, \mathcal{R}_{ab}
\end{equation}
The contraction between first and third indices of local Riemann tensor leads 
to Gauss-Codacci equation and a further contraction between the indices of 
Ricci tensor leads to,
\begin{equation}
 \mathcal{R} = 2(\kappa \rho + \sigma^2 - \omega^2) - \Theta^2 \frac{(n-1)}{n} \label{frw}
\end{equation}
where $\kappa = 8\pi G$, $2\,\sigma^2 = \sigma^{\alpha \beta}\sigma_{\alpha \beta}$ and 
$2\,\omega^2 = \omega^{\alpha \beta}\omega_{\alpha \beta}$. In the next section we consider the impact of 
extra dimensions on the evolution of shear and hence on the expansion scalar.

\section{Evolution Equations}
We consider the spacetime metric  which has three normal spatial 
dimensions and $D$ extra spatial dimensions, in addition to one time 
dimension \cite{isha}. The corresponding line element is given by
\begin{equation}
 ds^2 = -dt^2 + a(t)^2 \delta_{ij} dx^i dx^j + b(t)^2 \delta_{IJ} dX^I dX^J \label{flrwextra}
\end{equation}
where $i,j$ denotes $1,2,3$ and $I,J$ represents $4,5,...(D+3)$, extra spatial 
dimensions. Here, $D$ is a parameter which represents the number of extra dimensions and
takes integral values and hence, $n=3+D$. In the 
above metric, $a(t)$ denotes the scale factor in the normal dimensions and 
$b(t)$ represents the scale factor in the extra dimensions. We consider the 
whole $1+3+D$ dimensional Universe to be homogeneous and hence, $a(t)$ and 
$b(t)$ are functions only of time. The visible Universe is satisfactorily described 
by flat space i.e. spatial curvature is zero. For simplicity, we also 
assume extra-dimensional subspace to be flat.
For the line element given in equation (\ref{flrwextra}), spatial Ricci tensor $\mathcal{R}\,=\,0$ and 
$\omega=0$. 
Thus, equation (\ref{frw}) reduces to
\begin{equation}{}
 \kappa \rho + \sigma^2 = \Theta^2 \frac{(n-1)}{2n} \label{Sab}
\end{equation}
The expansion scalar corresponding to the metric (\ref{flrwextra}) is given as 
\begin{eqnarray}
 \Theta = 3\frac{\dot a}{a} \, + \, D \frac{\dot b}{b} \label{theta}
\end{eqnarray}
We follow the approach used in the paper \cite{caceres} to solve the governing 
equations  and to calculate the evolution of $\Theta$. The measure of total volume of the 
whole higher dimensional Universe is given as $S(t)^{n} = a(t)^3 \, b(t)^{D}$.

The solutions for the scale factors may be obtained directly from the 
Einstein equations as below \cite{ellis1999}: 
\begin{eqnarray}
 a(t) &=& S(t) \, \exp(\Sigma_1 W(t)), \nonumber \\
 b(t) &=& S(t) \, \exp(\Sigma_2 W(t))  \label{ab}
\end{eqnarray}
where $W(t)$ is defined as
\begin{equation}
  W(t) = \int \frac{dt}{S^{n}} \label{W(t)}
\end{equation}
and the constants $\Sigma_1$ and $\Sigma_2$  satisfy the relation 
$3 \Sigma_1 + D \Sigma_2=0$. 

Using the definition of shear, its components can be calculated to be
\begin{equation}
 \sigma_{ij} = \frac{D a^2}{n \, S^{n}}(\Sigma_1-\Sigma_2) \delta_{ij} 
\end{equation}
and using the relation $3 \Sigma_1 = - D \Sigma_2$, we get
\begin{equation}
 \sigma_{ij}= \frac{\Sigma_1 a^2}{S^{n}} \, \delta_{ij} \label{sigma11} 
\end{equation}
where $i,j \, = \, 1,2,3$. 
Similarly, we obtain 
\begin{equation}
 \sigma_{IJ}= \frac{\Sigma_2 b^2}{S^{n}}\, \delta_{IJ} \label{sigma44}
\end{equation}
where $I,J \, = \, 4,5,..., D+3$.
All off-diagonal components of shear tensor are zero. The scalar square 
magnitude of the shear, $\sigma^2$ defined 
by $\sigma^2=\sigma_{\alpha \beta} \sigma^{\alpha \beta}$ then becomes
\begin{equation}
 \sigma^2 = \frac{\Sigma^2}{S^{2n}} \label{sigma2n},
\end{equation}
where, 
\begin{equation}
 \Sigma^2=\frac{3\Sigma_1^2+D\Sigma_2^2}{2}
\end{equation}
here $\Sigma$ is a constant. Further, the expansion scalar $\Theta$ given by 
equation (\ref{theta}) can be written in terms of $S(t)$ as
\begin{equation}
 \Theta = n \,\frac{\dot{S}}{S} \label{thetan}
\end{equation}
Using equations (\ref{Sab}, \ref{sigma2n}, \ref{thetan}), one can obtain
\begin{equation}
 \frac{\dot{S}^2}{S^2} \frac{n \, (n-1)}{2} = \kappa \rho + \frac{\Sigma^2}{S^{2n}} \label{Seq}
\end{equation}
This equation constraints the flow of the fluid. We have to solve equation 
(\ref{Seq}) together with equation (\ref{W(t)}) for a  specific value of 
$n$ and specific type of density ($\rho$). By substituting the solutions back in 
equation (\ref{ab}), we can obtain $a(t)$ and $b(t)$. One then can study the 
evolution of shear components from  equations 
(\ref{sigma11}) and (\ref{sigma44}). Using the above mentioned expressions for the expansion scalar and 
the shear components, the Raychaudhuri equation  (\ref{rayeqn}) is manifestly 
satisfied. 

The energy momentum tensor of a general imperfect fluid can be expressed 
as~\cite{tsagas} 
\begin{equation}
 T_{\alpha \beta} = \rho u_{\alpha} u_{\beta} + p_a \delta_{\alpha}^i \delta_{\beta}^j h_{ij} + p_b \delta_{\alpha}^I \delta_{\beta}^J h_{IJ} + \pi_{\alpha \beta}  
\end{equation}
where $\rho$ is the total matter energy density, $p_a= w_a \rho$ is the 
pressure in normal dimensions and $p_b= w_b \rho$ is the pressure in extra 
dimensions. The tensor, $\pi_{\alpha \beta}$ is the symmetric and trace free anisotropic 
stress tensor as given by
\begin{equation}
 \pi_{\alpha \beta} = h_{(\alpha}^{\gamma} h_{\beta)}^{\delta} T_{\gamma \delta} - \frac{1}{n} h^{\gamma \delta} T_{\gamma \delta} h_{\alpha \beta}
\end{equation}
The conservation equation for total energy momentum tensor, expressed as $T^{\alpha \beta}_{;\beta}=0$ reduces to
\begin{equation}
 \dot{\rho} = -\frac{3 \dot{a}}{a} (\rho + p_a) - D \frac{\dot{b}}{b} (\rho + p_b) - \sigma^{\alpha \beta} \pi_{\alpha \beta} 
\end{equation}
which may further be re-written in terms of new variables as 
\begin{equation}
 \dot{\rho} = -\rho \, n \, \frac{\dot{S}}{S} + 3 \left(\frac{\dot{S}}{S} + \frac{\Sigma_1}{S^{n}}\right) \left(-p_a + \frac{(p_b -p_a)D}{n}\right) +  D \left(\frac{\dot{S}}{S} + \frac{\Sigma_2}{S^{n}}\right) \left(-p_b + \frac{3(p_a -p_b)}{n}\right) \label{rhof}.
\end{equation}
For simplicity, we assume that our universe is filled with radiation in visible subspace and 
with dust in extra-dimensional subspace. Thus, we have
$p_a \, = \, \frac{\rho}{3}$ and $p_b \, = \, 0$.

\section{Evolution of Expansion Scalar and Shear Components}
We may now numerically solve the three 
coupled first order differential equations (\ref{W(t)}), (\ref{Seq}) and 
(\ref{rhof}) to see the evolution of shear components in time 
and also the evolution of expansion scalar as given by the equation 
(\ref{thetan}). There may exist a variety of solutions depending on the
values of the independent model parameters, i.e., $D$ and $\Sigma_1$.
Before we do so, it is convenient to rescale the variables in terms of 
dimensionless quantities, namely
\begin{equation}
\begin{array}{rcl rcl rcl}
t &\equiv& \displaystyle \frac{\tau}{M}
& \qquad \quad & 
\bar{\rho} &\equiv& \displaystyle \frac{\rho}{M^4}  \label{rhoc}
& \quad & 
\Sigma_1 \rightarrow \displaystyle \frac{\Sigma_1}{M^4} \ ,
\end{array}
\end{equation}
where $M$ is the scale of quantum gravity in the entire bulk, and 
is related to $M_{\rm Planck}$ (a derived quantity defined
only for the theory in the $4$-dimensional subspace) through $V_D
M^{D+2}\, = \, M^2_{\rm Planck}$. Here $V_D$ is volume of the
extra-dimensional subspace.  One would expect that $V_D \gtrsim
M^{-D}$ and, thus $M \lesssim M_{\rm Planck}$.

%
%
%
%

\begin{figure*}
\centering
\begin{tabular}{cc}
\epsfig{file=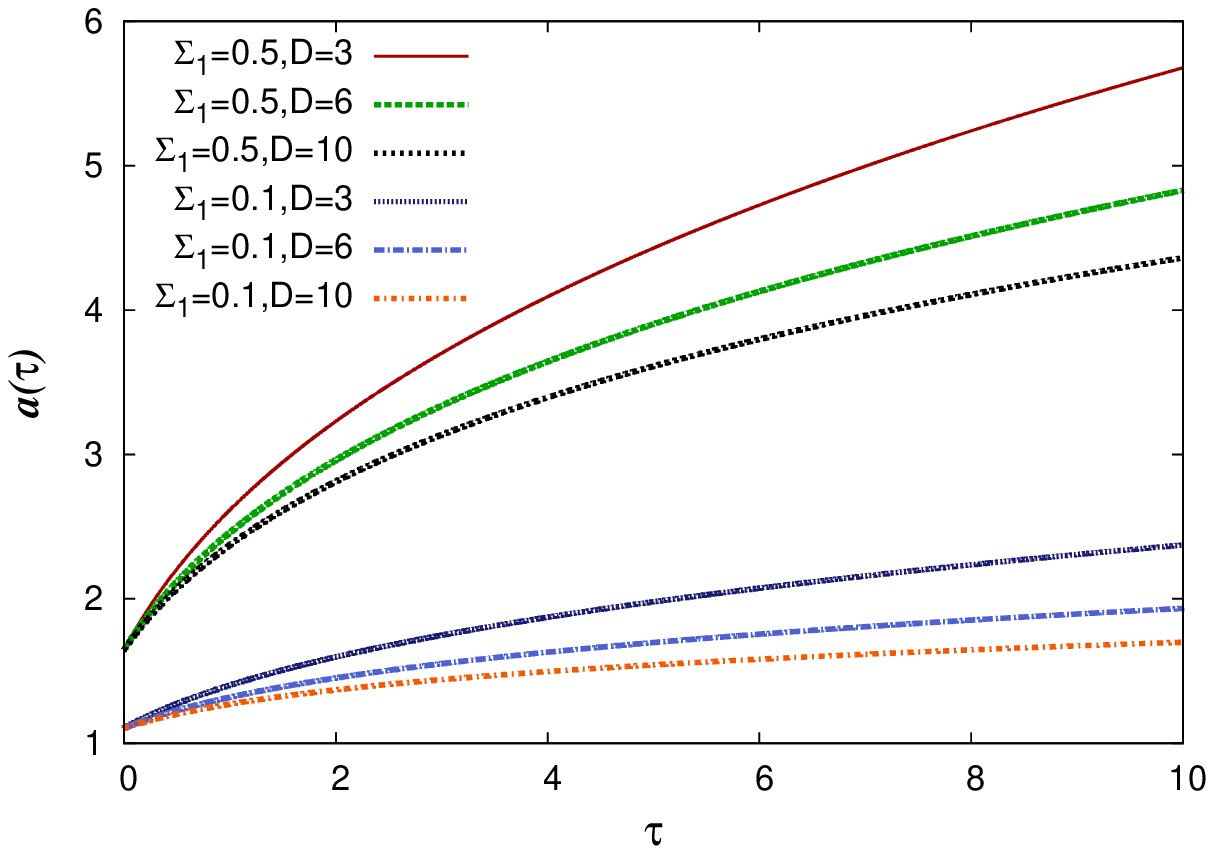,width=0.45\linewidth,clip=} & \epsfig{file=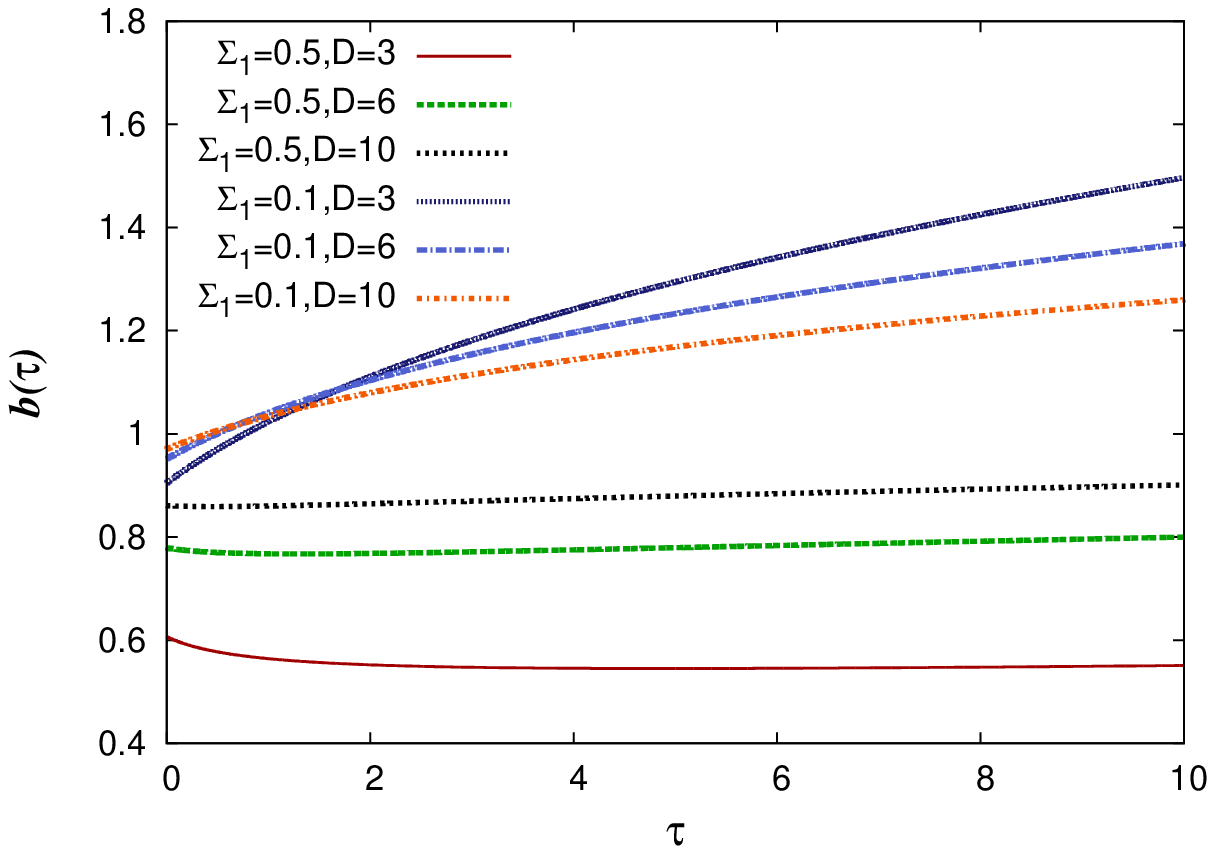,width=0.45\linewidth,clip=}\\
\end{tabular}
\caption{Evolution of scale factors $a(\tau)$ and $b(\tau)$ in the rescaled 
time for different values $\Sigma_1$ and $D$.}
\label{fig:scalef}
\end{figure*}
The equations are solved in terms of rescaled time.
The evolution of scale factors $a(\tau)$ and $b(\tau)$ in the
rescaled time for different values of $\Sigma_1$ and $D$ are shown in the figure 
\ref{fig:scalef}. It is interesting to see that the expansion in scale factor $a(\tau)$
is more rapid if we increase the amount of anisotropy parameter, $\Sigma_1$ in visible
subspace. This implies that shear helps in expansion. For the same $\Sigma_1$,
the amount of expansion of scale factor $a(\tau)$ is less by increasing the 
number of extra dimensions. It implies that increase in number of extra dimensions has a role
such that it reduces the effect of shear. For the case shown in figure \ref{fig:scalef}, the
scale factor in extra dimensional subspace flips its behavior, i.e., it is going from
contracting to expanding phase if we decrease $\Sigma_1$
but it becomes constant asymptotically. 
\begin{figure*}
\centering
\begin{tabular}{cc}
\epsfig{file=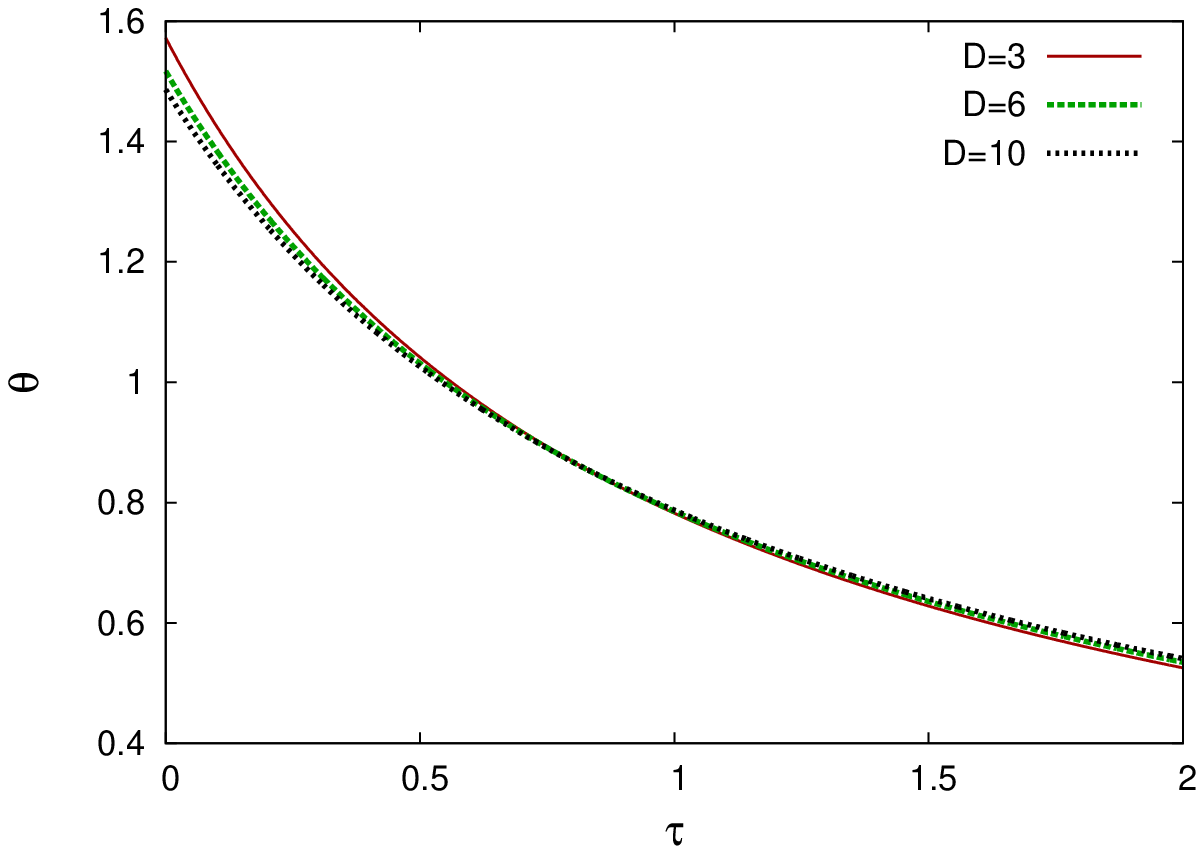,width=0.45\linewidth,clip=} & \epsfig{file=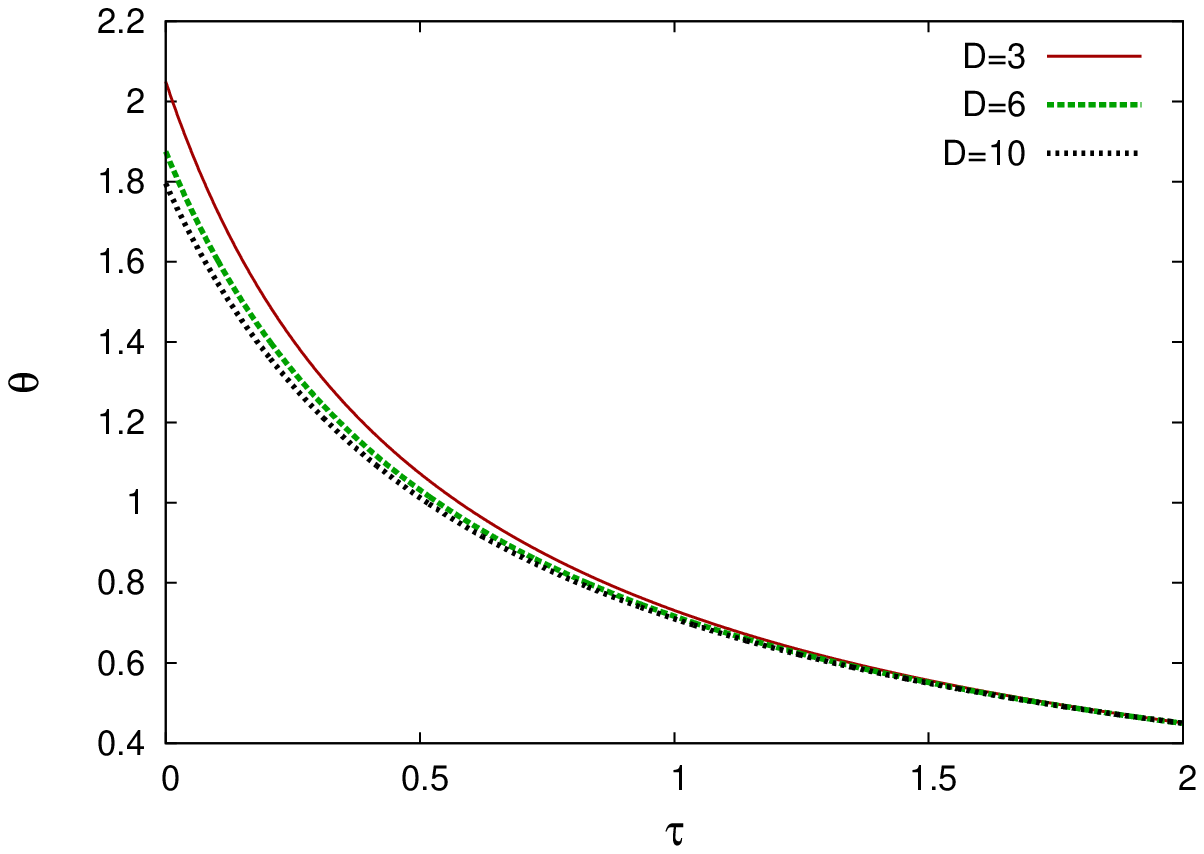,width=0.45\linewidth,clip=}\\
\end{tabular}
\caption{Evolution of the $\Theta$ 
 with respect to the rescaled time $\tau$ for $\Sigma_1 = 0.1$ (left panel) and 
 $\Sigma_1 = 0.5$ (right panel).}
 \label{fig:theta}
\end{figure*}

In figure \ref{fig:theta}, we present the evolution of $\Theta$ with respect to 
rescaled time for 
different number of extra dimensions, viz., $D=3,6,10$. Behaviour of $\Theta$ 
shows variation in the initial phase with the change in number of dimensions 
but as time evolves, it becomes independent of $D$ with a slight dependence on
 the value of $\Sigma_1$. 
\begin{figure*}
\centering
\begin{tabular}{cc}
\epsfig{file=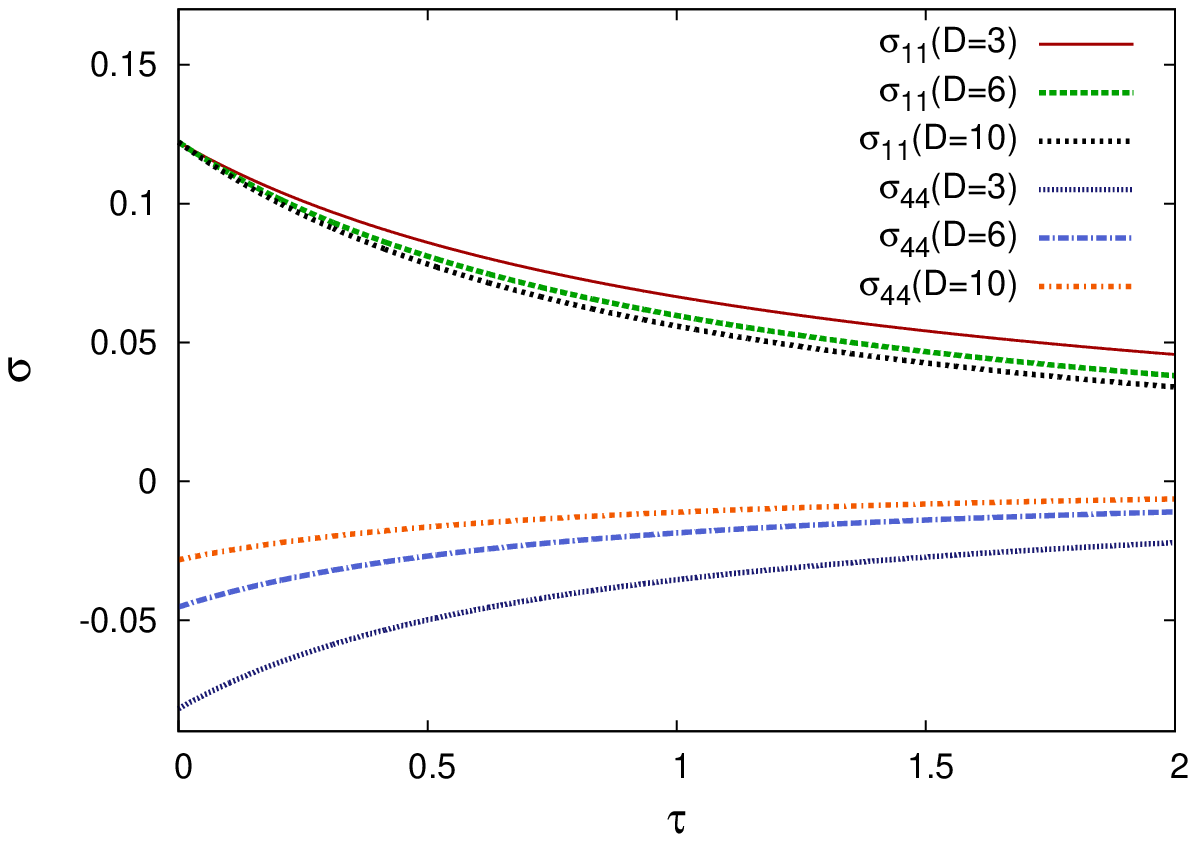,width=0.45\linewidth,clip=} & \epsfig{file=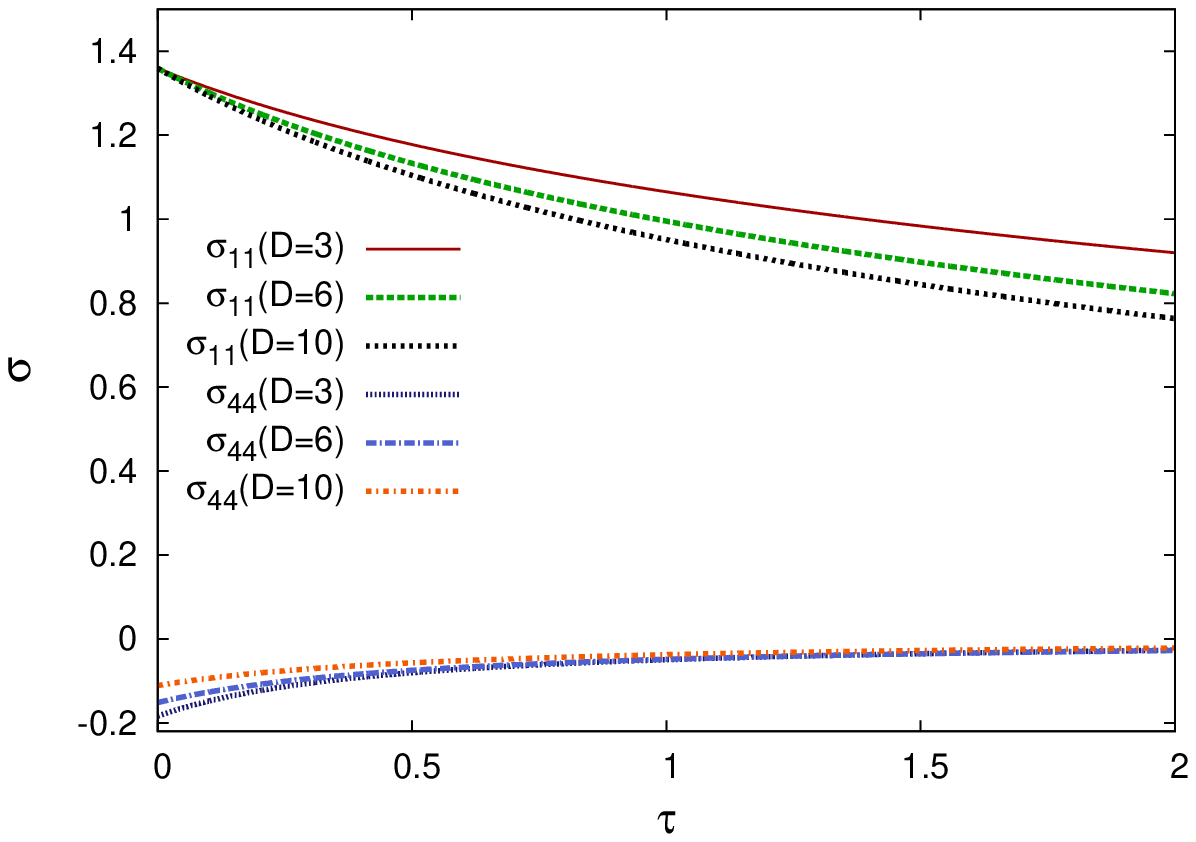,width=0.45\linewidth,clip=}\\
\end{tabular}
\caption{Behaviour of the shear components, $\sigma_{11}$ and $\sigma_{44}$ 
with the rescaled time $\tau$ for $\Sigma_1 = 0.1$ (left panel) and 
$\Sigma_1 = 0.5$ (right panel).}
\label{fig:sigma}
\end{figure*}
Figure \ref{fig:sigma} shows the variation of shear components as a 
function of time, where $\sigma_{11}$ and $\sigma_{44}$ represent shear in 
visible subspace and extra-dimensional subspace respectively. It is seen that
shear components in both subspace, i.e., visible and extra-dimensional subspaces
go to zero asymptotically. Higher the value of anisotropy parameter, larger the
time it takes to go to zero. With increase in extra dimensions, the value of shear 
components decreases and it tends to zero comparatively early.

\section{Conclusion}

In this work, we have studied the dynamical evolutions of expansion scalar and
shear in the higher dimensional FLRW Universe by considering that it is filled 
with the non-perfect fluid, which exerts different pressure in the normal and 
extra spatial dimensions. We have developed the dynamical equations governing
evolutions of expansion scalar and shear components involving the normal and
extra spatial dimensions. Our numerial study for different number of extra 
spatial dimensions with different values of independent parameter $\Sigma_1$ show 
that the extra dimensions have a very small effect
on the expansion scalar 
during some initial period of time and as time evolves, the expansion scalar 
become independent of number of extra spatial dimensions.

A number of interesting questions can arise from this mechanism which we 
plan to investigate in future. We plan to look at the effect of shear on
nucleosynthesis in this model as there will be a change in the 
expansion depending on $D$ and shear. One may get quite different results than the FLRW models.
Consequently, one can use the nucleosynthesis observations to limit
the shear constant, i.e. the anisotropy parameter $\Sigma_1$ and the number of
extra dimensions, $D$.    
We intend to report on this issue in near future.

{\bf Acknowledgements} The authors are grateful to Debajyoti Choudhury 
and T.R. Seshadri for helpful discussions and valuable suggestions. 
IP acknowledges the CSIR, India for assistance under grant
 09/045(0908)/2009-EMR-I and the facilities provided
  by the Inter University Center For Astronomy and Astrophysics, Pune,
  India through the IUCAA Resource Center (IRC), University of Delhi, New Delhi, India.
HN is thankful to the Department of Science and Technology (DST), 
New Delhi for financial assistance through SR/FTP/PS-31/2009.

\bibliographystyle{ws-ijmpd}
\bibliography{references}

\begin{thebibliography}{10}

\bibitem{Kaluza}
T.~{Kaluza}, {\em Sitz. Preuss. Akad. Wiss. Phys. Math.} {\bf K1}  (1921)
  966.

\bibitem{Klein}
O.~{Klein}, {\em Zeits. Phys.} {\bf 37}  (1926)   895.

\bibitem{tasi}
C.~Csaki  (2004) 605, \href{http://arxiv.org/abs/hep-ph/0404096}{{\ttfamily
  arXiv:hep-ph/0404096 [hep-ph]}}.

\bibitem{ADD}
N.~Arkani-Hamed, S.~Dimopoulos and G.~Dvali, {\em Phys.Today} {\bf 55N2}
  (2002) 35.

\bibitem{RS1}
L.~{Randall} and R.~{Sundrum}, {\em Physical Review Letters} {\bf 83} (Oct
  1999) 3370, \href{http://arxiv.org/abs/hep-ph/9905221}{{\ttfamily
  hep-ph/9905221}}.

\bibitem{RS2}
L.~{Randall} and R.~{Sundrum}, {\em Physical Review Letters} {\bf 83} (December
  1999) 4690, \href{http://arxiv.org/abs/hep-th/9906064}{{\ttfamily
  hep-th/9906064}}.

\bibitem{mattsson}
M.~Mattsson and T.~Mattsson, {\em JCAP} {\bf 1010}  (2010)   021,
  \href{http://arxiv.org/abs/1007.2939}{{\ttfamily arXiv:1007.2939
  [astro-ph.CO]}}.

\bibitem{ellis1999}
G.~F.~R. {Ellis} and H.~{van Elst}, { {Cosmological Models (Carg{\`e}se
  lectures 1998)}}, in {\em NATO ASIC Proc. 541: Theoretical and Observational
  Cosmology\/},  ed. M.~{Lachi{\`e}ze-Rey} (1999), pp. 1--116.

\bibitem{tsagas}
C.~G. {Tsagas}, A.~{Challinor} and R.~{Maartens}, {\em Physrep} {\bf 465}
  (August 2008) 61, \href{http://arxiv.org/abs/0705.4397}{{\ttfamily
  arXiv:0705.4397}}.

\bibitem{dadhich}
N.~{Dadhich}, {\em ArXiv General Relativity and Quantum Cosmology e-prints}
  (November 2005) \href{http://arxiv.org/abs/gr-qc/0511123}{{\ttfamily
  gr-qc/0511123}}.

\bibitem{ellis2007}
G.~F.~R. {Ellis}, {\em Pramana} {\bf 69} (July 2007)  ~15.

\bibitem{kar2007}
S.~{Kar} and S.~{Sengupta}, {\em Pramana} {\bf 69} (July 2007)  ~49,
  \href{http://arxiv.org/abs/gr-qc/0611123}{{\ttfamily gr-qc/0611123}}.

\bibitem{isha}
I.~{Pahwa}, D.~{Choudhury} and T.~R. {Seshadri}, {\em JCAP} {\bf 9} (September
  2011)  ~15, \href{http://arxiv.org/abs/1104.1925}{{\ttfamily arXiv:1104.1925
  [gr-qc]}}.

\bibitem{caceres}
D.~{Caceres}, L.~{Castaneda} and J.~M. {Tejeiro}, {\em ArXiv e-prints}  (March
  2010) \href{http://arxiv.org/abs/1003.3491}{{\ttfamily arXiv:1003.3491
  [gr-qc]}}.

\end{thebibliography}
\end{document}